\documentclass[a4paper,fleqn,usenatbib,useAMS]{mnras}

\usepackage{graphicx}	
\usepackage{amsmath}	
\usepackage{amssymb}	
\usepackage{multicol}        
\usepackage{bm}		
\usepackage{pdflscape}	
\usepackage{natbib}
\usepackage{times}
\usepackage{calc}
\usepackage{rotating}
\usepackage{lscape}
\bibpunct{(}{)}{;}{a}{}{,} 
\usepackage{longtable}
\usepackage{enumitem} 
\usepackage{hyperref}
\usepackage{pifont}
\usepackage{tikz}
\usepackage{xcolor}

\def\oversim#1#2{\lower0.5pt\vbox{\baselineskip0pt \lineskip-0.5pt
     \ialign{$\mathsurround0pt #1\hfil##\hfil$\crcr#2\crcr\sim\crcr}}}

\newcommand{\upsub}[1]{\sb{\mathrm{#1}}}
\newcommand{\upsup}[1]{\sp{\mathrm{#1}}}

\begingroup\lccode`~=`_\lowercase{\endgroup\let~\upsub}
\begingroup\lccode`~=`^\lowercase{\endgroup\let~\upsup}

\AtBeginDocument{%
  \catcode`_=12 \catcode`^=12
  \mathcode`_="8000
  \mathcode`^="8000
}

\usepackage[T1]{fontenc}
\usepackage{ae,aecompl}

\usepackage{mathptmx}

\def\aap {{A\&A}}

\def\apjl {{ApJ}}

\def\apjs {{ApJs}}

\title[ALMA polarization of OH 231.8+4.2]{ALMA reveals the coherence of the magnetic field geometry in OH 231.8+4.2}

\author[L. Sabin,et al.] 
{L. Sabin$^{1}$\thanks{E-mail:lsabin@astro.unam.mx (LS)}, R. Sahai$^{2}$, W.H.T. Vlemmings$^{3}$, Q. Zhang$^{4}$, A.A. Zijlstra$^{5}$, T. Gledhill$^{6}$,
\newauthor M. Huarte-Espinosa$^{7}$, A.F. P\'erez S\'anchez$^{8,9}$, E. Lagadec$^{10}$ and S.G. Navarro$^{11}$\\
$^{1}$Instituto de Astronom\'ia, Universidad Nacional Aut\'onoma de M\'exico, Apdo. Postal 877, C.P. 22860, Ensenada, B.C., M\'exico\\
$^{2}$Jet Propulsion Laboratory, MS 183-900, California Institute of Technology, Pasadena, CA 91109, USA\\
$^{3}$Department of Space, Earth and Environment, Chalmers University of Technology, SE-43992 Onsala, Sweden\\
$^{4}$Center for Astrophysics | Harvard \& Smithsonian, 60 Garden Street, Cambridge, MA 02138, USA \\
$^{5}$Jodrell Bank Centre for Astrophysics, School of Physics and Astronomy, University of Manchester, Oxford Road, Manchester M13 9PL, UK\\
$^{6}$School of Physics, Astronomy \& Mathematics, University of Hertfordshire, College Lane, Hatfield, Hertfordshire AL10 9AB, UK \\
$^{7}$Hewlett Packard Enterprise Data Science Institute, University of Houston, 4718 Calhoun Rd.
Houston, TX, 77204, USA.\\
$^{8}$European Southern Observatory, Alonso de C\'ordova 3107, Vitacura, Casilla 19001, Santiago de Chile, Chile \\
$^{9}$Leiden Observatory, Leiden University, Niels Bohrweg 2, 2333 CA Leiden, The Netherlands  \\
$^{10}$Universit\'e C$\hat{o}$te d'Azur, OCA, CNRS, Lagrange, France \\
$^{11}$ Instituto de Astronom\'ia y Meteorolog\'ia, Universidad de Guadalajara, 44130 Guadalajara, Jalisco, M\'exico }

\date{Accepted 2020 May 18. Received 2020 May 16; in original form 2020 April 13 }
\pubyear{2019}

\begin{document}

\label{firstpage}
\pagerange{\pageref{firstpage}--\pageref{lastpage}}
\maketitle


\begin{abstract}
In a continuing effort to investigate the role of magnetic fields in evolved low and intermediate mass stars (principally regarding the shaping of their envelopes), we present new ALMA high resolution polarization data obtained for the nebula OH 231.8+4.2. We found that the polarized emission likely  arises from aligned grains in the presence of magnetic fields rather than radiative alignment and self scattering.
The ALMA data show well organized electric field orientations in most of the nebula and the inferred magnetic field vectors (rotated by 90 degrees) trace an hourglass morphology centred on the central system of the nebula. One region in the southern part of OH 231.8+4.2 shows a less organized distribution probably due to the shocked environment. These findings, in conjunction with earlier investigations (maser studies and dust emission analysis at other scales and wavelengths) suggest an overall magnetic hourglass located inside a toroidal field. We propose the idea that the magnetic field structure is closely related to the architecture of a magnetic tower and that the outflows were therefore magnetically launched. While the current dynamical effect of the fields might be weak in the equatorial plane principally due to the evolution of the envelope, it would still be affecting the outflows. In that regard, the measurement of the magnetic field at the stellar surface, which is still missing, combined with a full MHD treatment are required to better understand and constrain the events occurring in OH 231.8+4.2.  \\

\end{abstract}


\section{Introduction}

The detection of magnetic fields in evolved low and intermediate--mass stars ($\sim$0.8--8 M\sun), such as AGB and post-AGB stars (pAGBs), pre-planetary (pPNe) and planetary nebulae (PNe), has generated great interest over the last decade. The main reason lies in our desire to understand the role magnetic fields could play in the dramatic departures from spherical symmetry occurring during the transition from the AGB to the PNe evolutionary phases \citep{Sahai2007,Sahai2011}.\\ 
Observations, detections and measurements of the field topology and strength have been performed using different methods: (spectro)polarimetry of various maser species (e.g. \citealt{Vlemmings2008,Ferreira2013,GOMEZ2016,Vlemmings2017}), measurement of circular polarization (Stokes {\it V}) in stellar atmospheres and photospheres (e.g. \citealt{Konstantinova2010, Lebre2014, Sabin2015a}), analysis of the linearly polarized emission of circumstellar dust and molecules (e.g. \citealt{Vlemmings2012,Vlemmings2017,Girart2012,Sabin2014}).
Nevertheless, we are still far from getting a full picture of the effects of magnetic fields in evolved stars. Small samples (compared to the total number of stars at each evolutionary stage) and few global analyses (combining the different techniques aforementioned), are likely the main causes for this. For instance, only two pAGBs are known to show a clear Zeeman splitting profile, and hence allow a definite magnetic field detection at their surface \citep{Sabin2015a}. This number plummets to zero when it comes to the central stars of PNe \citep{Leone2011,Jordan2012,Leone2014,Steffen2014}. \\
In this context the post-AGB OH 231.8+4.2 (a.k.a the Calabash nebula) establishes itself as a key object of study based on the various polarimetric observations that have been performed along the years. Indeed, this is one of the few objects for which numerous maser and dust polarization analysis have been undertaken, allowing for a unique view of its magnetic field.\\

\indent OH 231.8+4.2 is a bipolar oxygen-rich binary pAGB and has been the subject of maser and dust continuum polarization analyses targeting both the strength and topology of the magnetic field at different scales. \citet{Etoka2009} mapped the distribution of OH masers at 1667 MHz with MERLIN, at an angular resolution of $\sim$0.2 \arcsec and found a ring-like pattern with a $\sim 4$\arcsec\, diameter centred on the central star. This corresponds to a radius of $\sim$3300 AU considering that OH 231.8+4.2 is part of the open cluster M46 (or NGC 2347, \citet{Jura1985}) for which the recent Gaia DR2 release indicates a mean parallax of 0.61 mas \citep{Cantat2018}, leading to a distance to the nebula of 1639 pc. The polarization analysis (circular and linear) indicated a radial magnetic field aligned with the outflow of the pAGB.\\
Later on, \citet{Leal2012} observed the H$_{2}$O 6$_{1,6}$--5$_{2,3}$ rotational transition with the Very Long baseline Array (VLBA) and a synthesized beam size of $\sim$1.7$\times$0.9 mas. Of all the 30 detected maser spots, most of them are distributed in the direction of the collimated outflow and at $\sim$40 AU from the central star (assuming a distance of 1540 pc derived  by \citealt{Choi2012}), only a few show detectable level of polarization. Although it was not possible to estimate the direction of the magnetic field via linear polarization analysis, the presence of circular polarization in two masers spots located at opposite sides of the star indicated a strength of the magnetic field along the line of sight of 44$\pm$7 mG and -29$\pm$21 mG. The extrapolation of the field's strength to the stellar surface, assuming a toroidal distribution, sets a value of $\sim$2.5 G at 1 R$\star$.\\
Finally, \citet{Sabin2014,Sabin2015b} mapped the polarized thermal emission of the dust continuum at submillimeter (0.87mm/345 GHz) and millimeter (1.3mm/230 GHz) range with the Submillimeter Array (SMA) and Combined Array for Research in Millimeter-wave Astronomy (CARMA), respectively. Based on the radiative torques (RATs) theory regarding the alignment of non-spherical spinning dust grains with respect to the magnetic field \citep{Lazarian2007,Lazarian2008,Hoang2008,Lazarian2011}, the authors were able to infer the presence of an X-shaped magnetic field distribution centred at the base of each optical outflow and encompassing the red- and blue-shifted $^{12}$CO ({\it J}=3$\rightarrow$2) molecular outflows. The 230 GHz polarimetric analysis also indicates an organized magnetic field, which not only mirrors part of the X-shaped structure seen at the submillimeter wavelength, but most importantly it also shows the presence of a new structure aligned with the equatorial plane of the nebula. The combined analysis thus revealed the coexistence of an inner dipole/polar magnetic field configuration and an outer toroidal configuration. In addition, the analysis of the variation of ``magnetic vectors'' (hereafter, $\overrightarrow{B}$-vectors) position angles (PA) led the authors to also consider the possibility of an helical magnetic field geometry. All the detected features seem to point towards a magnetic collimation and launching mechanism in OH 231.8+4.2. \\ 
Thus, in this article we present new high resolution polarimetric observations (at 0.5\arcsec) obtained with the Atacama Large Millimeter/submillimeter Array (ALMA). These data provide new insights into the inner dusty region of OH 231.8+4.2 by allowing us to probe the magnetic engine (assuming a dynamo mechanism) closer to its launch-site and complete a multi-scale analysis of the magnetic field in OH 231.8+4.2. We present the ALMA observing procedure in \S\ref{obs} and the results are shown in \S\ref{cont}. A comparative analysis with all the previously obtained data is presented in \S\ref{comp} \& \S\ref{comp2}. Our concluding remarks are shown in \S\ref{final}.

\begin{figure}
\includegraphics[height=7cm]{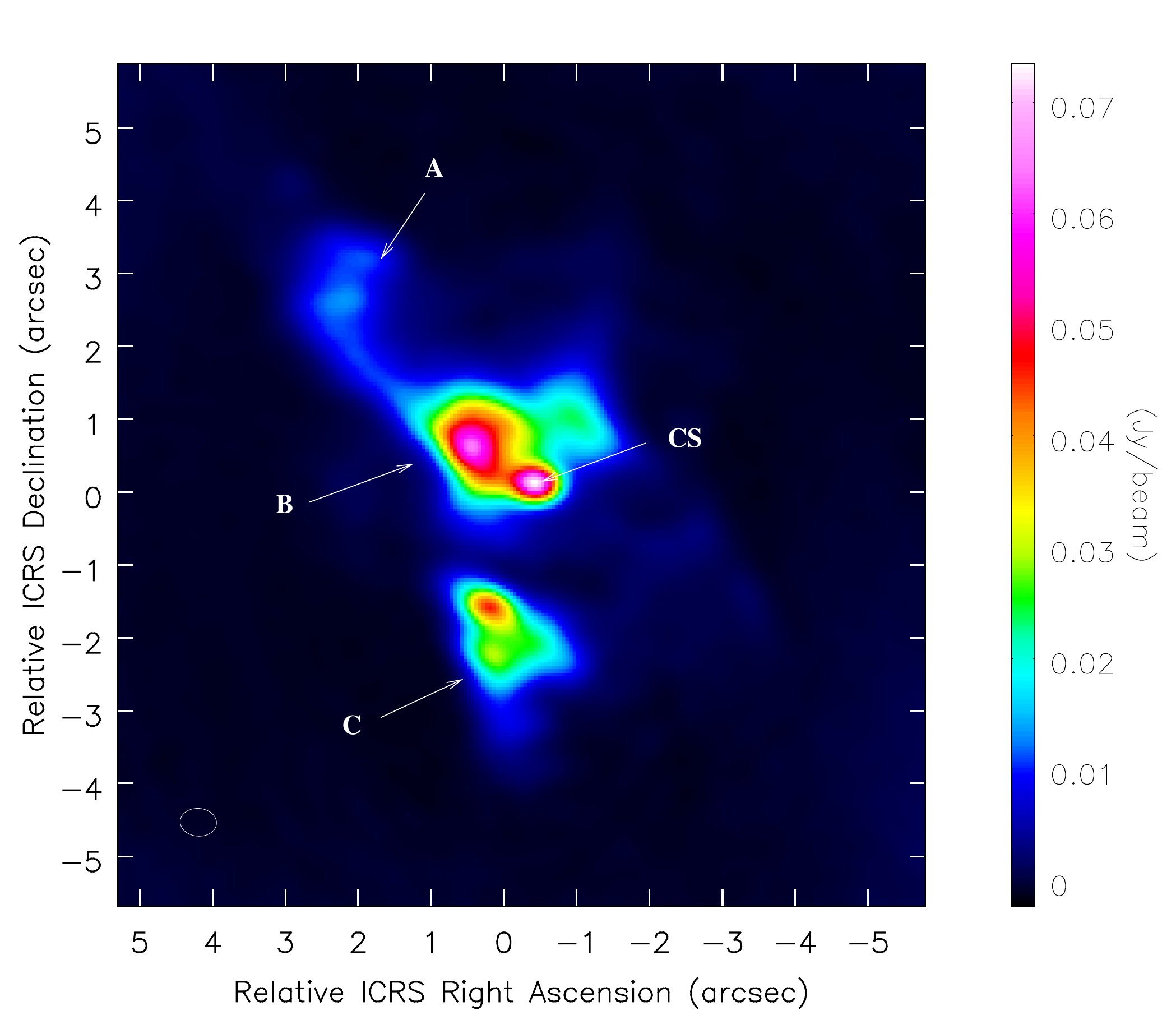}
\caption{ALMA dust continuum emission map, at 345 GHz, of OH 231.8+4.2. The global hourglass shape is well defined and the peak emission is coincident with the central star location (CS on the map). The origin (0,0) of the map corresponds to the coordinates $\alpha(J2000)$=$07^{h}\,42^{m}\,16^{s}\,947$, $\delta(J2000)=-14\degr\,42\arcmin\,50\farcs199$. North is on top and east is on the left, the synthesized beam is indicated at the lower left corner.}
\label{dust}
\end{figure}

\begin{figure}
\includegraphics[height=7.9cm]{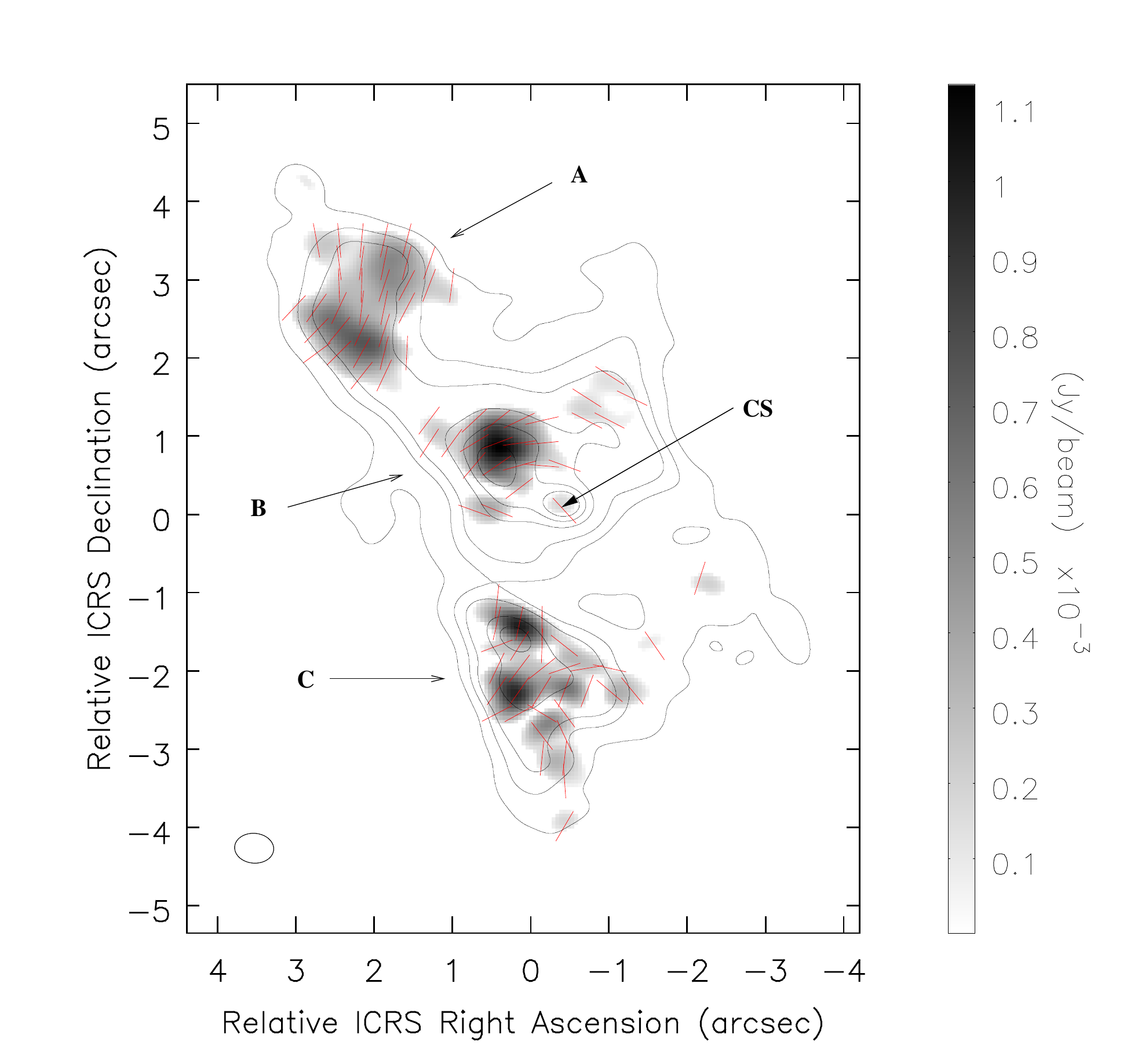}
\caption{ALMA polarization map (at 345 GHz) showing the electric distribution of vectors. The contours are drawn in steps of [0.01,0.05,0.1,0.3,0.5,0.7]$\times$73.6 mJy beam$^{-1}$. The letters A,B and C indicate the three main polarized areas. The peak polarized emission, with a value of 1.13$\pm$0.13 mJy beam$^{-1}$, is found in the region B. The central star location is also indicated (CS on the map). North is on top and east is on the left, the synthesized beam is indicated on the left corner.}
\label{Evectors}
\end{figure}

\begin{figure*}
\includegraphics[height=8.5cm]{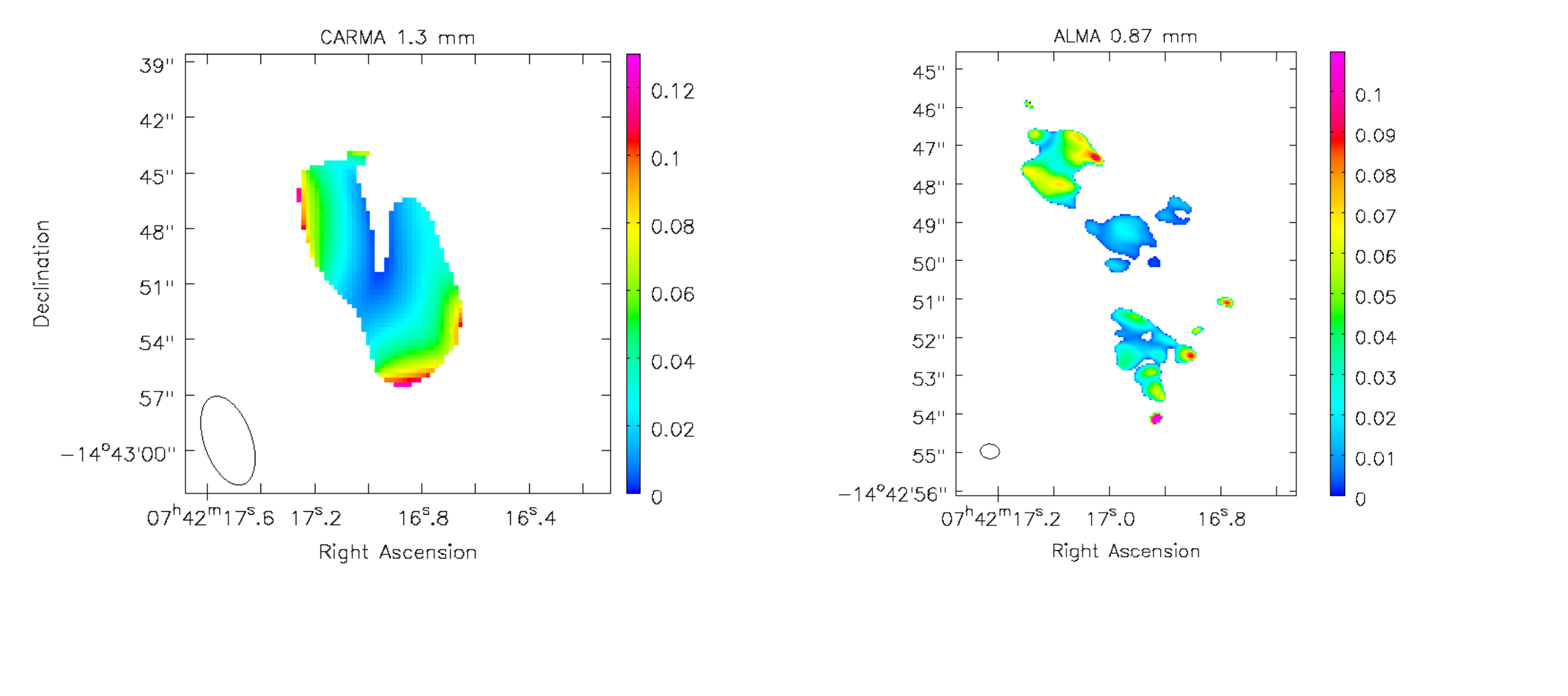}
\vspace{-1cm}
\caption{Distribution of the degree of polarization in the CARMA (left) and ALMA data (right). The wedge indicates the percentage polarization. }
\label{Percentpol}
\end{figure*}

\section[]{Observations and data reduction}\label{obs}

The non-standard polarimetric observations were performed with 40$\times$12m ALMA antennae on 2016-12-28. The observations, under the code 2016.1.00196.S (P.I. Sahai), were done using frequency Band 7 with spectral windows centred at 336.495 GHz, 338.432 GHz, 348.495 GHz and 350.495 GHz, each window having 1.875 GHz bandwidth and 64 channels. The C40-3 configuration was used with a maximum baseline of 460m and the maximum recoverable scale (MRS) was 3.9\arcsec.
Two datasets were taken, with a total time of 58.22 min and 91.55 min respectively (including calibrations such as bandpass, flux, phase and polarization) and OH231.8+4.2 was observed for 30.23 min in both cases. The quasars which were used for calibration purposes are J0750+1231 as the flux and polarization calibrator, J0730-1141 as the phase calibrator, and J0522-3627 as the bandpass calibrator. All three quasars were also used for removal of effects of the atmosphere and pointing calibrations.\\

The {\it uv } data reduction was performed using the ALMA pipeline and imaging was done with The Common Astronomy Software Applications \citep[CASA v 4.7.2]{McMullin2007}. The two executions were calibrated separately, then combined before imaging and polarization calibration.
To achieve a better sensitivity and unveil more details, two rounds of self calibration were performed. We note that dynamical range of the Stokes {\it I} image is at best around 300 (despite the self-calibration and careful removal of the channels with line emission), indicating that all the flux is not recovered. With a maximum recoverable scale (MRS) of 3.9\arcsec we have resolved out all the smooth structures larger than this size. A Briggs weighting with a robust factor of 0.5 was used to generate continuum maps with a resulting synthesized beam of 0.50\arcsec$\times$0.38\arcsec with a position angle PA= 85.63$\degr$. The images were also primary-beam corrected and we note that the leakage is expected to be lower than 2\% and the 1$\sigma$ continuum instrumental error is around 0.1\%. \\
The resulting Stokes {\it I} (intensity) image has an rms noise level $\sigma_{I}\approx$ 0.260 mJy beam$^{-1}$ and the Stokes {\it Q} and {\it U} (polarized intensity) images have rms noise levels $\sigma_{Q}\approx\sigma_{U}\approx$ 0.021 mJy beam$^{-1}$. All the maps generated have a 3$\sigma$ cut.

\section[]{ALMA dust polarization analysis}\label{cont}

Fig.\ref{dust} presents the distribution of the dust continuum emission (Stokes {\it I}) toward the center of OH 231.8+4.2. The hourglass shape (also seen in Fig.\ref{Evectors}) is obvious and extends over $\sim$8\arcsec$\times$3\arcsec. The eastern side of the distribution appears to be brighter than the western counterpart. We observe two major bright regions noted B and C in Fig.\ref{dust} .
The first bright section (B) is confined at the waist of the hourglass structure, with the brighter part on its eastern side, and shows a mean flux of 29$\pm$3 mJy beam$^{-1}$ with a peak intensity (P$_{\it{Int}}$ for {Stokes \it I} emission) of 74$\pm$8 mJy beam$^{-1}$. The latter, located at the coordinates $\alpha(2000)$=07$^{h}$\,42$^{m}$\,16$^{s}$92,  $\delta(2000)=-14\degr\,42\arcmin\,50\farcs076$ is coincident with the position of the central star component QX PuP derived by \citet{Dodson2018} and \citet{SC2018} ($\alpha   (J2000)$=07$^{h}$\,42$^{m}$\,16$^{s}$915, $\delta (J2000)= -14\degr\,42\arcmin\,50\farcs06$). The other region (C) is located in the south-east with a mean flux of 23$\pm$3 mJy beam$^{-1}$ and a peak of 46$\pm$5 mJy beam$^{-1}$. \\

The polarization map, shown in Fig.\,\ref{Evectors}, indicates the global distribution of the electric vectors ($\overrightarrow{E}$-vectors) and we emphasize that the dusty regions in the nebula where those vectors are not indicated have a polarized intensity below 3$\sigma$.
The polarized emission is not uniform but is rather patchy with three main regions (noted A, B and C in Fig.\,\ref{Evectors}). The peak polarization (P$_{\it{Pol}}$), located at the coordinates 07$^{h}$\,42$^{m}$\,16$^{s}$976\, -14$\degr$\,42$\arcmin$\,49$\farcs$357 inside the area B, has a measured flux of 1.13$\pm$0.3 mJy beam$^{-1}$. As observed in many astronomical objects, including OH 231.8+4.2 (see \citealt{Sabin2014,Sabin2015b}), P$_{\it{Pol}}$ and P$_{\it{Int}}$ are not spatially coincident. \\
Most of the $\overrightarrow{E}$-vectors are relatively well organized within each area and this is particularly seen in regions A and B. 
We found mean position angles for the electric vectors (EVPA) of about -19\degr\,$\pm$2\degr\, and -42\degr$\pm$3\degr\, for the regions A and B respectively. 
However, while the mean PA seems to indicate a global preferential alignment direction of the dust grains in these two regions, locally we observe a curvature in the distribution of the position angle of the $\overrightarrow{E}$-vectors, which is likely to be of interest for the understanding of the magnetic field.
In region C, the orientations of the $\overrightarrow{E}$-vectors do not appear to show a coherent pattern, making their interpretation more difficult.
Finally, we note that the peak fractional polarization decreases from the outer to the central regions. Its value ranges from 9.7\% and 16.4\% in the (outer) regions A and C respectively down to 2.2\% in the (inner) region B.

 \begin{table}
\caption[]{\label{tabcomp} Summary of the available dust polarimetric observations.}
\begin{tabular}{|c|c|c|c|}
\hline 
Array & CARMA & SMA & ALMA \\ 
\hline 
\hline 
$\nu$ (GHz) & 230 & 345 & 345 \\ 
\hline 
Beam Size ($\arcsec$) & 4.6$\times$2.4 & 2.5$\times$1.9 & 0.5$\times$0.4\\ 
\hline 
$\Theta{\arcsec}$ & 15.3$\times$8.8 & 7.3$\times$5.7 & 8.0$\times$3.0 \\ 
\hline 
$D_{pc}$ & 0.15$\times$0.08& 0.07$\times$0.05  &0.08$\times$0.03  \\ 
\hline 
$D_{AU}$ & 30600$\times$17600& 14600$\times$11400 & 16000$\times$6000  \\ 
\hline 

\end{tabular}
\begin{minipage}{8.5cm}
\hspace{-1cm}
\hspace{1cm} $\Theta{\arcsec}$, $D_{pc}$ and $D_{AU}$ indicate the angular extent and approximate size of the thermal emission in arcsec, parsec and AU  respectively, assuming a distance of 1639 pc for the nebula. The two last values are corrected from the 35\degr\,inclination angle of the source.
\end{minipage}
\end{table}

\begin{figure*}
\hspace{-0.8cm}
\includegraphics[height=10cm]{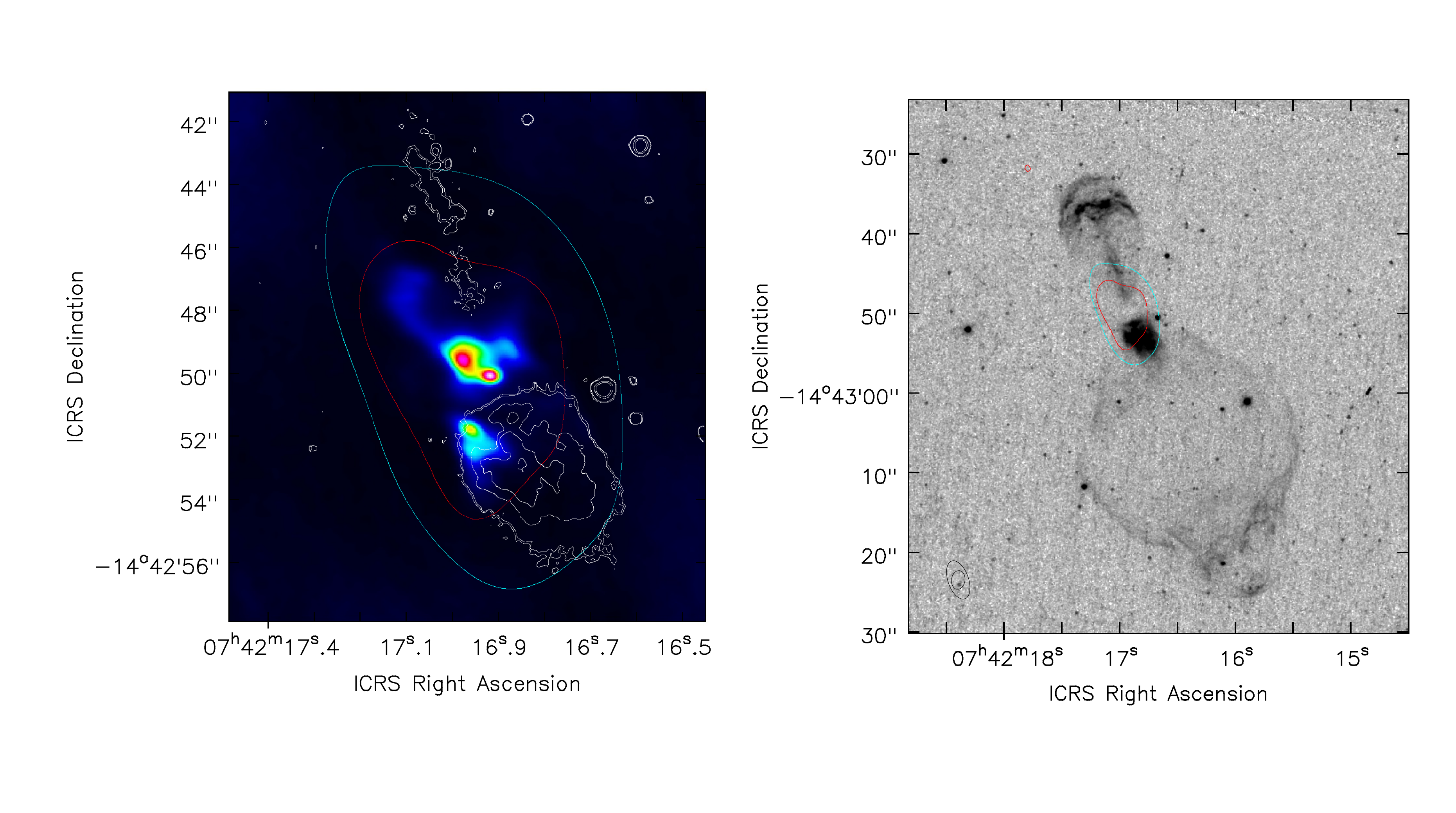}
\vspace{-1cm}
\caption{(Left) The panel shows the lowest significant contours of the dust continuum emission distribution observed with CARMA (blue, contour level of 0.005 Jy beam$^{-1}$) and SMA (red, contour level of 0.03 Jy beam$^{-1}$) superimposed on an  ALMA image (in color scales) of the continuum emission at 345 GHz. The white contours show the optical emission from the source obtained with the {\it Hubble Space Telescope} and WFPC2 at 0.65$\mu$m (F656N filter). (Right) A full view of the bipolar nebula OH231.8+4.2 as seen with the HST in H$\alpha$ emission (F656N filter), together with the same dust emission contours as shown in the left panel. }
\label{area}
\end{figure*}

\begin{figure*}
\hspace{-0.8cm}
\includegraphics[height=9cm]{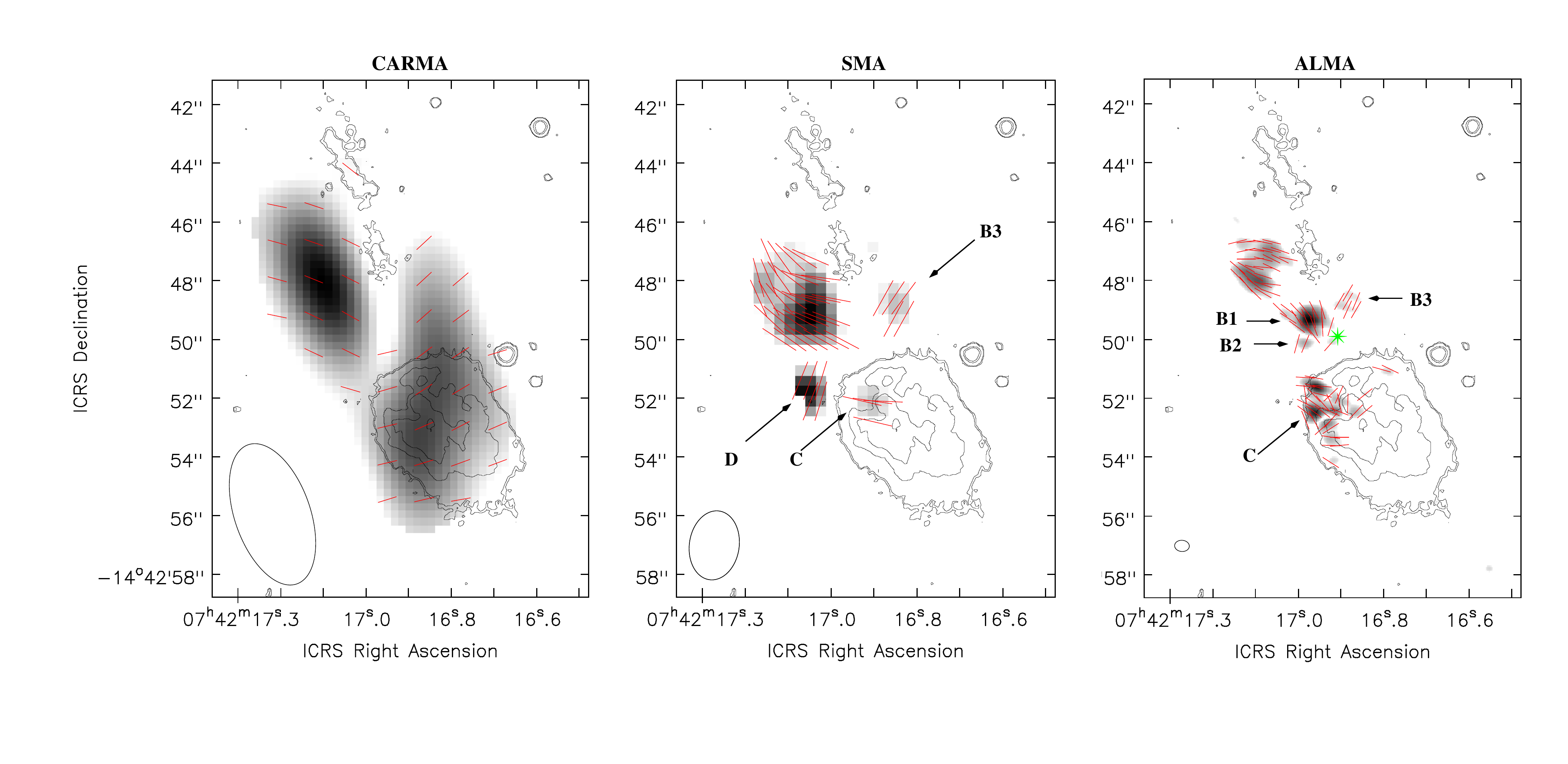}
\vspace{-1cm}
\caption{Polarized dust emission maps (greyscale) with the E-vectors (red) rotated by 90\degr\,to illustrate the magnetic field geometry obtained with CARMA (left), the SMA (center) and ALMA (right) in each case the beam is shown on the bottom left corner. The green asterisk in the right panel indicates the location of the central star and for all the maps we  superimposed the contours of the optical emission of the source (obtained with the HST). Note that the length of the vectors are not related to the percentage polarization.}
\label{CompB}
\end{figure*} 

\begin{figure}
\includegraphics[height=8.5cm]{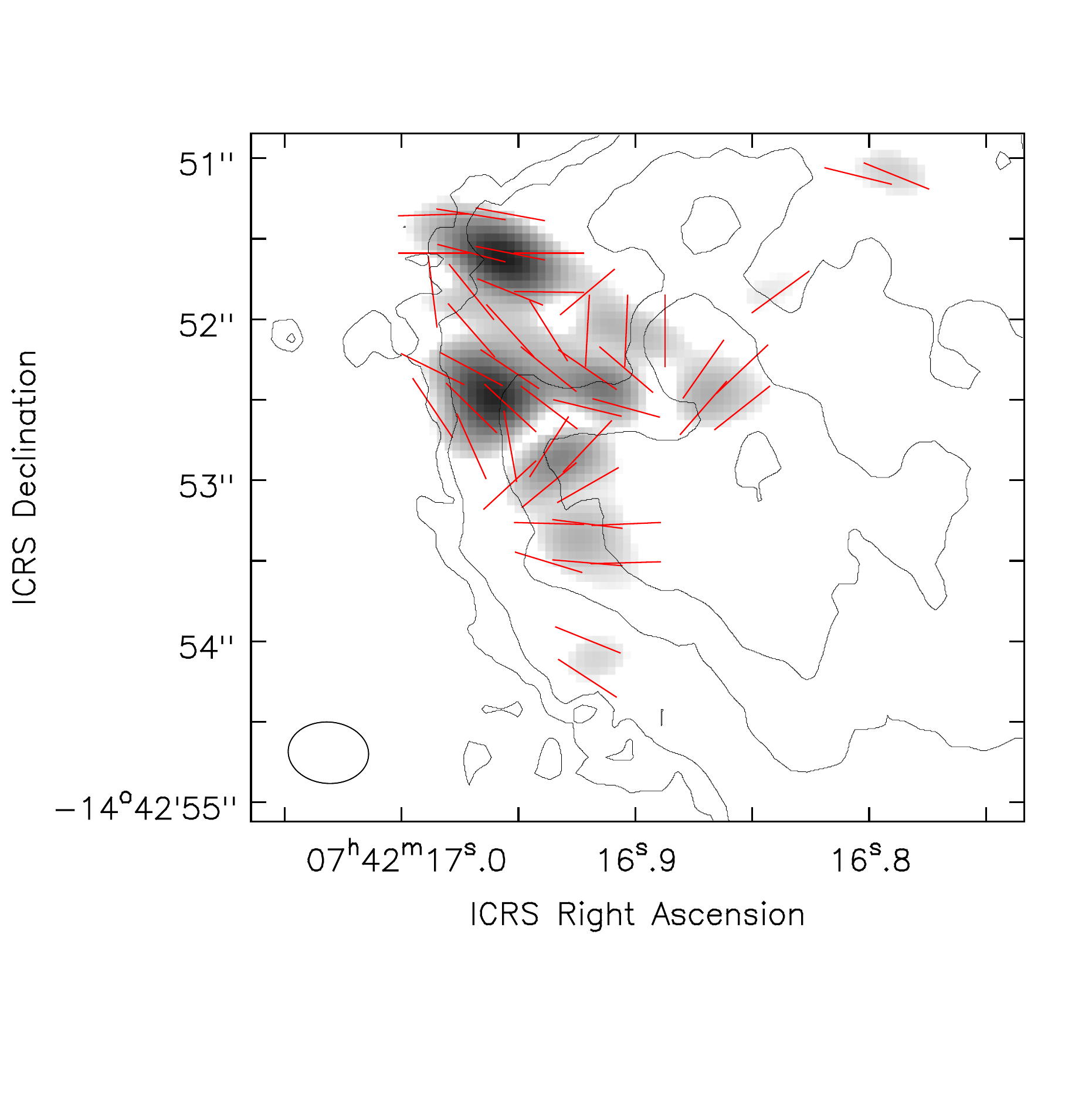}
\vspace{-1cm}
\caption{ALMA map of the magnetic field distribution of vectors generated with a zoom on the region ''C''. The beam size is shown on the left lower corner. }
\label{ZoomC}
\end{figure}

\section[]{Multi-scale dust polarization analysis}\label{comp}

It is generally assumed that the polarization of ISM thermal dust emission observed at long  (e.g. submillimeter and millimeter) wavelengths is due to spinning non-spherical dust grains aligned with respect to the magnetic field. However, two other polarization mechanisms have been proposed, mostly in the study of protoplanetary discs. \\
First, mm-sized grains may not be aligned by the magnetic field (due to the slow Larmor precession), but rather by the radiative flux, as shown by \citet{Tazaki2017}. In this case, the polarization traces the radiation anisotropy (see figure 12 of \citealt{Tazaki2017}). 
The application of this mechanism to our bipolar evolved nebula might produce different results due to the different physical and morphological conditions in OH231.8+4.2 compared to protoplanetary disks. However, based on the general distribution of the polarization vectors indicated by the disc model in the case of radiative polarization, we can make a first guess on whether this process could explain the polarization seen in OH 231.8+4.1. In the case of grain alignment caused by a radiative flux the polarization distribution would more likely be centro-symmetric (or spherical symmetric). 
The millimeter data obtained with CARMA indicates that only the $\overrightarrow{E}$-vectors located in the upper area, corresponding to the regions A and B in our new ALMA dataset, are consistent with radiative grain alignment. However, the high resolution submillimeter ALMA data (Fig.\,\ref{Evectors}) show an arc-like distribution of the polarization vectors in region A (see their position angles) which deviate from the spherical symmetric pattern around the central star. This implies that radiative grain alignment (or alignment with radiation anisotropy) is not the dominant mechanism for grain alignment here. 
It is possible that the polarization in the inner region of the nebula (area B, see Fig.\ref{Evectors}) is due to the radiation field, since many of the $\overrightarrow{E}$-vectors in this area could be related to a spherical symmetric geometry. However, this pattern is incomplete as it is not seen in the southern part of region B, and the mean percentage polarization is relatively low ( $\simeq$1\%), both at 230 GHz (CARMA data) and at 345 GHz (ALMA data). So we disregard radiative alignment as the main polarization process operating in OH 231.8+4.2. One would expect higher percentage at this location, or in other words, a more efficient alignment of the dust grains.  
Detailed modelling, involving different radiation field intensities, observing wavelengths, and dust grain sizes (\citealt{Ohashi2018, Tazaki2017}), would be needed to assess if and in which conditions this alignment process would be viable.\\
The second polarization mechanism unrelated to magnetic fields is due to dust self-scattering. This process becomes important if the dust grain size is comparable to the wavelength, i.e. if a$_{max}$ $\sim$ $\lambda$/$2\pi$ with a$_{max}$ the maximum grain size \citep{Kataoka2015} and has been mostly studied at millimeter wavelengths. 
Hence, a low P\% is achieved when a$_{max}$ $>$ $\lambda$/$2\pi$ and vice versa. Fig.\,\ref{Percentpol} indicates the variation of the degree of polarization in the nebula at millimeter and sub-millimeter wavelengths. Although there is a clear gradient, which can be associated with a depolarization effect, P\% is generally always $\geq$ 2\% in most of the nebula except for the region closer to the central star. The relatively high degree of polarization that we find in OH231.8+4.2, compared to the maximum value typically expected from models of self-scattering, about 2.5\% at 345 GHz (a$_{max}$=100 $\mu$m) \citep{Kataoka2015}, suggest that self-scattering is not the dominant polarization factor in OH 231.8+4.2. The polarization process occurring in the inner region is less clear though. \\
In addition, the good agreement between the 230 GHz and 345 GHz maps in terms of polarization vectors distribution indicates that there is no wavelength dependence. We further convolved the ALMA Stokes {\it Q} and {\it U} images to the CARMA resolution and confirmed this trend. As scattering polarization is frequency dependent, it is very unlikely that self-scattering is a main element in the polarization seen in the nebula.
It is important to note that the different polarization processes are not mutually exclusive.
If, as we have argued above, the radiation flux (via radiation anisotropy alignment) and self-scattering are not principal factors operating in the nebula, magnetic fields are likely responsible for the dust alignment and the observed polarization.\\

Hence using the ALMA data and bearing in mind the above caveats, it is therefore possible to proceed with the mapping of the ``magnetic architecture'' of OH 231.8+4.2.  Moreover, combining our new findings with the CARMA and SMA data, we now have multi-scale information on the dust distribution, and by extension on the magnetic field geometry. Table \ref{tabcomp} compiles the characteristics of each available dataset and Fig.\,\ref{area} shows the extent of the thermal dust emission for the different arrays as well as the HST H$\alpha$ emission. It is therefore possible to pinpoint the distribution of the thermal dust emission with respect to the ionized emission. Viewed on the plane of the sky, the former is centred on the central star of the nebula and extends at most only over a semi-major axis of $\sim$7\arcsec. This represents $\sim$2.6 times less than the optical northern lobe and 5.1 times less than the southern one. The thermal dust emission observed with our combined dataset, therefore traces the highest dust density regions of the nebula (see also \citealt{Balick2017} and \citet{SC2018}).
Fig.\,\ref{CompB} presents three distinct behaviours (or distributions) of the magnetic fields in the regions above and below the equatorial plane. They are described in the following sections \ref{reg1}, \ref{reg2} and \ref{reg3}.

\subsection[]{Poloidal or hourglass magnetic distribution}\label{reg1}

The analysis of the ALMA polarization data shows that the $\overrightarrow{B}$-vectors (EVPA rotated by 90\degr\,) located in the areas A (with a mean PA=63\degr\,), B1 (with a mean PA=23\degr\,) and B3 (with mean a PA=-25\degr\,) align quite well with the edges of the northern outflow within its opening angle of 67\degr\, (with a mean PA= 23\degr\, based on the optical HST image presented here). The uncertainty on the polarization angles, assuming a 3$\sigma$ threshold, was estimated to be around 10\% using the relation derived by \citet{Serkowski74}.
We can now integrate all the information to assess the distribution of the magnetic field. With the equator of the nebula (at PA $\sim$113\degr\,) as a demarcation line, the northern section shows an organized ``$\overrightarrow{B}$-vectors'' V-pattern along the outflows in the 230 GHz CARMA dataset. As it gets closer to the central region, with an increase in resolution and change of wavelength (870 $\mu$m SMA and ALMA dataset), we observe a remarkable consistency in the general distribution of the vectors. This would suggest that the mechanism responsible for aligning the dust grains, that we assume to be the magnetic field, is strong and coherent enough (the vectors showing a noticeable organized pattern) that it is not disturbed along most of the extent of the northern lobe.\\ 
Another significant point is that from the global ``V-shaped'' structure observed with CARMA (Fig.\,\ref{CompB}-left), a complete hourglass becomes even clearer down to the ALMA resolution. Indeed, the set of $\overrightarrow{B}$-vectors in the region noted ``D'' in the SMA observations at PA=-20\degr\, (Fig.\,\ref{CompB}-middle) and the ones in the region noted ``B2'' in the ALMA observations at PA= -18\degr\, (Fig.\,\ref{CompB}-right) seems to draw the southern section of the magnetic hourglass. In all the maps, the center of this bipolar pattern appears to be coincident with the approximate location of the central star. \\

While the Stokes I flux densities of both the SMA and ALMA maps are roughly comparable with values of $\sim$1.65 Jy and $\sim$1.02 Jy respectively, some differences in the polarized emission distribution are worth mentioning. Hence, the strong submillimeter emission in the region ``D'' (seen with the SMA) is no longer visible in the ALMA data. In addition we observe other discrepancies in the polarization distribution between both maps. Indeed, the strong central polarization in the region A seems to be missing in the ALMA map and conversely the strong polarization in region B1 (seen with ALMA) is not present in the SMA map. We excluded the ''contamination'' of our previous SMA maps by some line polarization and conducted independent ALMA data re-reductions. The reason for the absence of structures seen with the SMA (at lower resolution) in the ALMA map (having a higher resolution) is not clear. 

\subsection[]{A large scale toroidal field component}\label{reg2}

The southern part of the magnetic hourglass structure observed with ALMA, and even the SMA, appears to be surrounded or embedded within a wide toroidal like pattern, with a size of $\sim$6\arcsec$\times$7\arcsec, located at larger scale. This can be seen through the orientation of the $\overrightarrow{B}$-vectors in the CARMA image \citet{Sabin2015b} (Fig.\ref{CompB}-left). This toroidal pattern totally disappears as we are getting closer to the central star perhaps due to the changes in the properties or geometry of the alignment mechanism. At first glance the magnetized torus, seen at 230 GHz, could be considered as an independent secondary element of the magnetic structure of OH 231.8+4.2 (along with the poloidal structure).\\
However, \citet{Sabin2015b} noted a coherent variation in the PA of the magnetic vectors observed in the millimeter range, between the northern and southern sections (with respect to the equatorial plane), which would be consistent with a single helical structure (composed of a toroidal and a poloidal sections). This will be discussed below.

\subsection[]{The case of the region ``C''}\label{reg3}

While the structures described in \S\S\,\ref{reg1} and \ref{reg2} show organized magnetic vectors and preferential directions, this is not the case for the region noted ``C'' in Figs. \ref{Evectors} and \ref{CompB}. A closer look at this area (Fig. \ref{ZoomC}) indicates that a coherent pattern seems to exist within each internal clump, however this local organization does not hold when the clumps are taken altogether. The changes in the polarization pattern could be linked to the  shocked environment at the base of the southern lobe. \citet{Balick2017} presented a detailed hydrodynamical model of OH 231.8+4.2 where they distinguished both lobes. The region coincident with our area ``C'' is represented with a faster ejection (400 km s$^{-1}$ against 225 km s$^{-1}$ for the northern lobe), a wider opening angle (50\degr / 30\degr) and a higher injection mass rate (2.3e$^{-2}$ M$_{\odot}$ yr$^{-1}$ / 4.8e$^{-3}$ M$_{\odot}$ yr$^{-1}$). These conditions would therefore contribute to the explanation for the lack of organization in the distribution of vectors.

\begin{figure*}
\includegraphics[height=8cm]{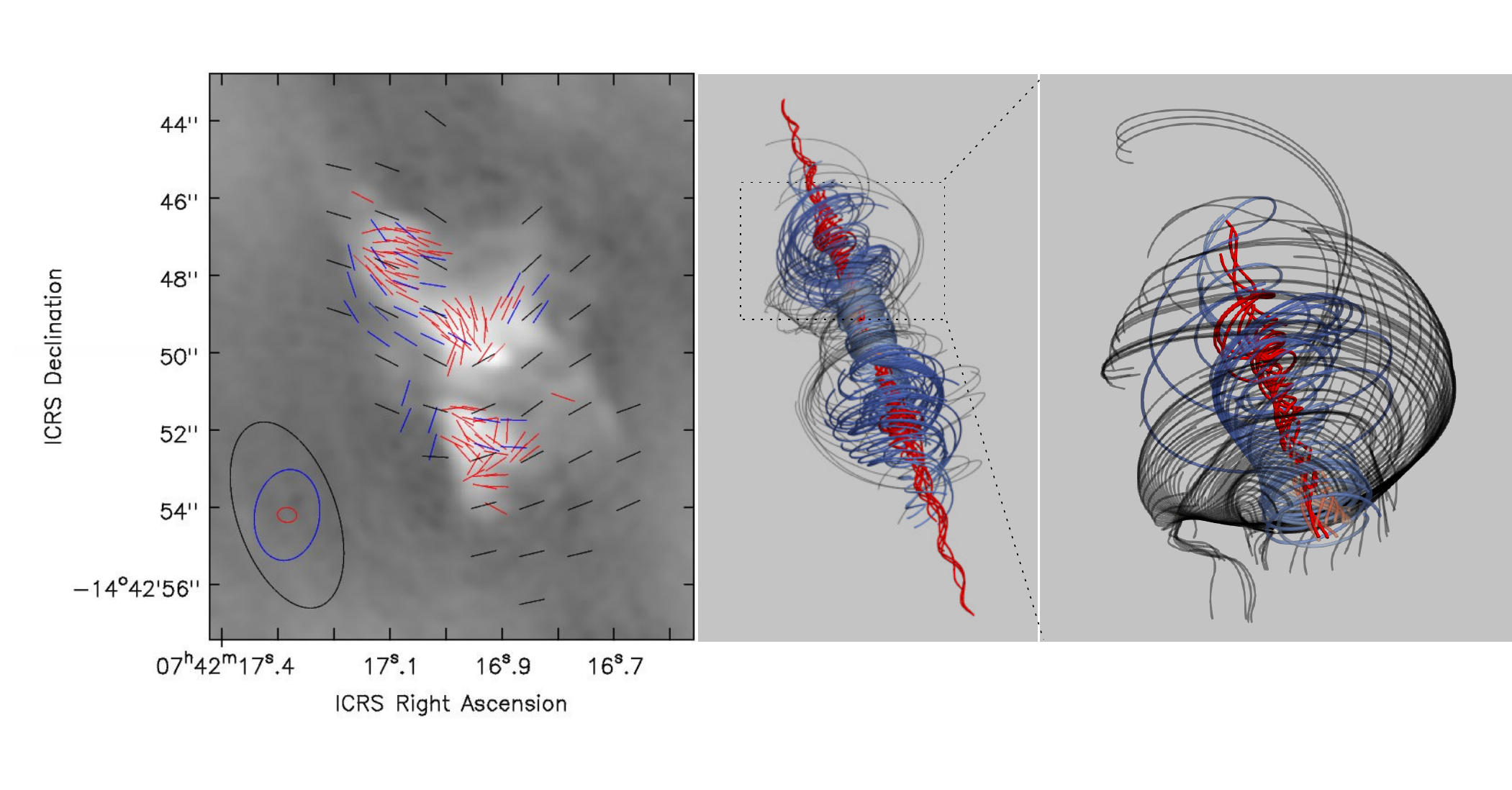}
\vspace{-1cm}
\caption{ The image on the left shows the distribution of the ``magnetic vectors'' taking into account all arrays: CARMA (black), SMA (Blue) and ALMA (Red). On the middle and right panels, sketches of the full magnetic tower as well as a zoom on the upper part of the tower are shown with the same color code to mirror the magnetic lines from each array. This could be a good representation of magnetic distribution in OH 231.8+4.2. }
\label{Tower}
\end{figure*}

\section[]{Global View of Magnetic Fields in OH 231.1+4.2}\label{comp2}
We can now use the new thermal dust polarization observations with existing OH maser data by \citet{Etoka2009} and \citet{Leal2012}, to trace the magnetic field pattern from $\sim$30600 AU down to $\sim$40 AU from the central star (and by extrapolation to its surface). 
\subsection[]{Overall magnetic topology}\label{topology}

The OH masers \citep{Etoka2009} and dust polarization emission data (SMA, ALMA) indicate the presence of a well organized magnetic field in the polar direction and in the "inner" parts of the nebula (i.e. the regions closer to the central star). The most spatially distant magnetic field lines seen with CARMA point towards the presence of an additional toroidal magnetic pattern. Such a dual distribution of the magnetic field is a strong reminiscence of magnetically driven outflows model \citep{Huarte2012} and is also supported by the possible presence of an helical field, inferred by the distribution of the polarization vectors (i.e. PA variation) in the 230 GHz polarization map (Fig.\ref{Tower}).\\

Now that we have established the presence of a multi-scale magnetic field in OH 231.8+4.2 we can postulate that it might have a dynamical role in the shaping of the nebula.\\
The polarization vectors seen at millimeter and sub-mm wavelengths could indicate the presence of frozen-in toroidal magnetic field lines which would have subsequently been dragged along with the outflows, but they could also trace the remnant of a dynamo-generated magnetic field which has evolved in time. In the later scheme the field is amplified in the rotating disk/torus or within the binary system. The smaller the disk or the closer the binary separation, the higher the magnetic field intensity, and therefore we would expect a magnetic launching mechanism. MHD launching is very likely the only process that would reproduce the kinetic energies observed in such evolved nebulae \citep{Garcia2020}. Hence, \cite{Huarte2012} (and references therein) discussed two magnetic models, which could account for the observed magnetic structure in OH 231.8+4.2: magnetic towers and magneto-centrifugal launch. The main distinction between the two models is the distance up to which the magnetic energy density dominates over the kinetic energy. Based on the magneto-hydrodynamics simulations realised by the authors, magneto-centrifugal launch may dominate below $\simeq$100 R$_{star}$ (inside the Alfv\'en radius). At larger distances the flow is mostly directed by hydrodynamical motions. Magnetic towers on the contrary can have an effective action at much larger scales. We present in Fig.\ref{Tower} a model of a magnetic tower, performed with 3D-MHD numerical simulations (see \citealt{Huarte2012}), with different spatial layers of magnetic field lines which seems to closely mirror the distribution seen in OH 231.8+4.2. The sketch represents the core of the tower only. Knowing the magnetic field intensity of the stellar engine is therefore a key element to determine first, if the field is strong enough to trigger the magnetic launch and then to discern which magnetic launching mechanism applies. The only available data are the extrapolation given by \citet{Leal2012} indicating a surface magnetic field of $\sim$2.5G. But as this value originates from a single H$_{2}$O maser emission located at $\sim$40 AU it cannot be fully established as representative of the stellar surface. \citet{Sabin2014} deduced that for B $\geqslant$ 0.6 mG, the magnetic pressure ($P_{B}=B^{2}/8\pi$) was dominating over the thermal pressure ($P_{th}=n_{H}kT$) in the central region ($\simeq$ 6\arcsec in size or 12000 AU) defined by \citet{Bujarrabal2002}. This corresponds to the area mapped by ALMA. Assuming a temperature of 100 K for a region much closer to the central star ($\simeq$0.5\arcsec or 1000 AU in size, see \citealt{SC2018}), the lower limit rises to 1 mG.\\
Based on the distances where the fields were observed and the strength derived, it is unlikely that a dynamo-like activity would still be at work in OH231.8+4.2 or at least that it would have any dynamical effect in terms of launching, but magnetic fields can still be involved in shaping of the outflows. \\

While the main morphology would be defined by the magnetic tower, there is still the
possibility that internal shocks or other localised hydrodynamical motions could locally
break the symmetry inside the nebula. For example, \citet{Balick2017} using purely hydrodynamical modeling, find that one way to
explain the asymmetry in the lobes is to assume that they are created by the interaction
of a fast, bipolar wind with an ambient circumstellar envelope, with the former having
different speeds, mass-injection rates and opening angles in the North and South. Presumably, a full MHD simulation that can reproduce the overall large asymmetry in the lobes would have to incorporate similar asymmetries in the magnetically-launched wind, thus implying an underlying asymmetry in the magnetic field at the base of the outflows.
Asymmetries in the overall dust distribution, as well as the unexplained structures that we see in the dust distribution (e.g. region C) could presumably represent regions of locally enhanced density and/or temperature, resulting from cooling and associated instabilities (e.g., see Fig. 2 of \citealt{Huarte2012}).\\
The Calabash nebula is somewhat peculiar because of the large striking difference in the size of its North and South lobes. However its dusty nature makes it an ideal target for the type of investigation we present in this article, which addresses
fundamental questions regarding the presence of large scale, organized magnetic fields in
pre-planetary nebulae in general, and their role in the formation and shaping of these
objects. We observed either toroidal or polar magnetic structures in other sources that have been
investigated in the past. OH 231.8+4.2 is the only object with a clear X-shaped morphology
(but also the best characterised). Although it is too early, and the data too sparse, for
us to comment on the commonality of such geometries, the observations presented in this
paper can enable testing of theoretical mechanisms involving magnetic fields for shaping
PPNe.

As aforementioned, a complete MHD model (i.e. taking into account the detailed hydrodynamics information of the source) would therefore greatly help in determining the field strength needed to trigger the magnetic structures that are observed and if/how it would affect the dynamics of the outflows. \\

\section[]{Summary and Conclusions}\label{final}

We present a polarization analysis of OH 231.8+4.2 with ALMA at the high spatial resolution of 0.5''. The new data  allowed us to have a clear picture of the magnetic field geometry at smaller scales. In combination with other interferometric polarization data, we now have new insights into the whole magnetic field structure which might govern or be closely related to the object geometry. The main findings are described below.\\

\noindent$\bullet$  ALMA revealed well organized $\overrightarrow{E}$-vectors in the bright A and B regions corresponding to the northern outflow and the waist of OH 231.8+4.2. The $\overrightarrow{B}$-vectors are consistent with an hourglass morphology whose origin is coincident with the location of the central star of the nebula. In the southern outflow, the polarization vectors (region C) are globally less aligned perhaps due to a shock dominated environment.\\
\noindent$\bullet$  The absence of a global spherical symmetric pattern, the high percentage polarization ($\geq$2\%) in most of the nebula and the absence of a wavelength dependence in terms of the polarization vector distribution argue against radiative alignment and self scattering as the main mechanisms involved in the dust grain alignment pattern, leaving the magnetic field as the most probable cause for the polarization observed. However the polarization process(es) at work in the inner region is less clear and would deserve further study.\\
$\bullet$  The combination of all the polarimetric data gathered for OH 231.8+4.2 allows us to perform a multi-scale analysis. The latter tends to confirm the presence (at all scales) of a  magnetically collimated outflow most likely originating from the area hosting the central star of the nebula (and where a dynamo would have been at work). This magnetic hourglass, present in the zones of highest dust density regions, is embedded in a toroidal pattern and the whole structure closely mirrors the architecture of a magnetic tower.  \\

While we probed the presence of a well organized magnetically collimated structure, the precise measurement of the magnetic field in the central engine is still missing. This piece of information is needed to accurately determine the actual (and past) dynamical effects of the field onto the shaping the nebula. While we could only be seeing the remnant of a fossil field dragged by the outflows, the hypothesis of the occurrence of a MHD launching at some early stage in the history of OH 231.8+4.2, which would still have an impact on the outflows geometry, should not be discarded and has been explored in this article.  \\

\section*{Acknowledgements}

The authors thank the referee for the careful review of this paper.
LS would like to thank the ALMA/NAASC Staff (in particular Erica Keller and Sarah Wood) for the support provided at the NA-ARC Headquarters at Charlottesville (US). LS also thanks Guillermo Garc\'ia Segura for the useful discussions about MHD launching as well as M.A.\,Guerrero for commenting on the paper. LS also acknowledges support from the Fundaci\'on Marcos Moshinsky. 
RS's contribution to the research described here was carried out at the Jet Propulsion Laboratory (JPL), California Institute of Technology, under a contract with NASA, and was funded in part by NASA through an Astrophysics Data Analysis Program award "An X-Ray and UV Study of a New Class of AGB stars with Actively-Accreting Binary Companions: fuv AGB Stars ? (17-ADAP17-0206) and grant number HST-AR-10681.01-A from the Space Telescope Science Institute (operated by AURA, Inc., under NASA contract NAS 5-26555). MHE acknowledges that this work was completed in part with resources provided by the Research Computing Data Core at the University of Houston.
ALMA is a partnership of ESO (representing its member states), NSF (USA) and NINS (Japan), together with NRC (Canada), MOST and ASIAA (Taiwan), and KASI (Republic of Korea), in cooperation with the Republic of Chile. The Joint ALMA Observatory is operated by ESO, AUI/NRAO and NAOJ.
The National Radio Astronomy Observatory is a facility of the National Science Foundation operated under cooperative agreement by Associated Universities, Inc.
\bibliographystyle{mn2e}

\bibliography{OH231}

\begin{thebibliography}{}

\bibitem[\protect\citeauthoryear{{Balick}, {Frank}, {Liu} \&
  {Huarte-Espinosa}}{{Balick} et~al.}{2017}]{Balick2017}
{Balick} B.,  {Frank} A.,  {Liu} B.,    {Huarte-Espinosa} M.,  2017, \apj, 843,
  108

\bibitem[\protect\citeauthoryear{{Bujarrabal}, {Alcolea}, {S{\'a}nchez
  Contreras} \& {Sahai}}{{Bujarrabal} et~al.}{2002}]{Bujarrabal2002}
{Bujarrabal} V.,  {Alcolea} J.,  {S{\'a}nchez Contreras} C.,    {Sahai} R.,
  2002, \aap, 389, 271

\bibitem[\protect\citeauthoryear{{Cantat-Gaudin}, {Jordi}, {Vallenari},
  {Bragaglia}, {Balaguer-N{\'u}{\~n}ez}, {Soubiran}, {Bossini}, {Moitinho},
  {Castro-Ginard}, {Krone-Martins}, {Casamiquela}, {Sordo} \&
  {Carrera}}{{Cantat-Gaudin} et~al.}{2018}]{Cantat2018}
{Cantat-Gaudin} T.,  {Jordi} C.,  {Vallenari} A.,  {Bragaglia} A.,
  {Balaguer-N{\'u}{\~n}ez} L.,  {Soubiran} C.,  {Bossini} D.,  {Moitinho} A.,
  {Castro-Ginard} A.,  {Krone-Martins} A.,  {Casamiquela} L.,  {Sordo} R.,
  {Carrera} R.,  2018, \aap, 618, A93

\bibitem[\protect\citeauthoryear{{Choi}, {Brunthaler}, {Menten} \&
  {Reid}}{{Choi} et~al.}{2012}]{Choi2012}
{Choi} Y.~K.,  {Brunthaler} A.,  {Menten} K.~M.,    {Reid} M.~J.,  2012, in
  {Booth} R.~S.,  {Vlemmings} W.~H.~T.,   {Humphreys} E.~M.~L.,  eds, Cosmic
  Masers - from OH to H0 Vol.~287 of IAU Symposium, {Trigonometric Parallax of
  the Protoplanetary Nebula OH 231.8+4.2}.
pp 407--410

\bibitem[\protect\citeauthoryear{{Dodson}, {Rioja}, {Bujarrabal}, {Kim}, {Cho},
  {Choi} \& {Youngjoo}}{{Dodson} et~al.}{2018}]{Dodson2018}
{Dodson} R.,  {Rioja} M.,  {Bujarrabal} V.,  {Kim} J.,  {Cho} S.~H.,  {Choi}
  Y.~K.,    {Youngjoo} Y.,  2018, \mnras, 476, 520

\bibitem[\protect\citeauthoryear{{Etoka}, {Zijlstra}, {Richards}, {Matsuura} \&
  {Lagadec}}{{Etoka} et~al.}{2009}]{Etoka2009}
{Etoka} S.,  {Zijlstra} A.,  {Richards} A.~M.,  {Matsuura} M.,    {Lagadec} E.,
   2009, in {Soonthornthum} B.,  {Komonjinda} S.,  {Cheng} K.~S.,   {Leung}
  K.~C.,  eds, The Eighth Pacific Rim Conference on Stellar Astrophysics: A
  Tribute to Kam-Ching Leung Vol.~404 of Astronomical Society of the Pacific
  Conference Series, {The Geometrical and Magnetic Structure of the
  Proto-Planetary Nebula OH 231.8+4.2 Traced by OH Maser Emission}.
p.~311

\bibitem[\protect\citeauthoryear{{Garc{\'\i}a-Segura}, {Taam} \&
  {Ricker}}{{Garc{\'\i}a-Segura} et~al.}{2020}]{Garcia2020}
{Garc{\'\i}a-Segura} G.,  {Taam} R.~E.,    {Ricker} P.~M.,  2020, arXiv
  e-prints, p. arXiv:2003.06073

\bibitem[\protect\citeauthoryear{{Girart}, {Patel}, {Vlemmings} \&
  {Rao}}{{Girart} et~al.}{2012}]{Girart2012}
{Girart} J.~M.,  {Patel} N.,  {Vlemmings} W.~H.~T.,    {Rao} R.,  2012, \apjl,
  751, L20

\bibitem[\protect\citeauthoryear{{G{\'o}mez}, {Uscanga}, {Green}, {Miranda},
  {Su{\'a}rez} \& {Bendjoya}}{{G{\'o}mez} et~al.}{2016}]{GOMEZ2016}
{G{\'o}mez} J.~F.,  {Uscanga} L.,  {Green} J.~A.,  {Miranda} L.~F.,
  {Su{\'a}rez} O.,    {Bendjoya} P.,  2016, \mnras, 461, 3259

\bibitem[\protect\citeauthoryear{{Hoang} \& {Lazarian}}{{Hoang} \&
  {Lazarian}}{2008}]{Hoang2008}
{Hoang} T.,  {Lazarian} A.,  2008, \mnras, 388, 117

\bibitem[\protect\citeauthoryear{{Huarte-Espinosa}, {Frank}, {Blackman},
  {Ciardi}, {Hartigan}, {Lebedev} \& {Chittenden}}{{Huarte-Espinosa}
  et~al.}{2012}]{Huarte2012}
{Huarte-Espinosa} M.,  {Frank} A.,  {Blackman} E.~G.,  {Ciardi} A.,  {Hartigan}
  P.,  {Lebedev} S.~V.,    {Chittenden} J.~P.,  2012, \apj, 757, 66

\bibitem[\protect\citeauthoryear{{Jordan}, {Bagnulo}, {Werner} \&
  {O'Toole}}{{Jordan} et~al.}{2012}]{Jordan2012}
{Jordan} S.,  {Bagnulo} S.,  {Werner} K.,    {O'Toole} S.~J.,  2012, \aap, 542,
  A64

\bibitem[\protect\citeauthoryear{{Jura} \& {Morris}}{{Jura} \&
  {Morris}}{1985}]{Jura1985}
{Jura} M.,  {Morris} M.,  1985, \apj, 292, 487

\bibitem[\protect\citeauthoryear{{Kataoka}, {Muto}, {Momose}, {Tsukagoshi},
  {Fukagawa}, {Shibai}, {Hanawa}, {Murakawa} \& {Dullemond}}{{Kataoka}
  et~al.}{2015}]{Kataoka2015}
{Kataoka} A.,  {Muto} T.,  {Momose} M.,  {Tsukagoshi} T.,  {Fukagawa} M.,
  {Shibai} H.,  {Hanawa} T.,  {Murakawa} K.,    {Dullemond} C.~P.,  2015, \apj,
  809, 78

\bibitem[\protect\citeauthoryear{{Konstantinova-Antova}, {Auri{\`e}re},
  {Charbonnel}, {Drake}, {Schr{\"o}der}, {Stateva}, {Alecian}, {Petit} \&
  {Cabanac}}{{Konstantinova-Antova} et~al.}{2010}]{Konstantinova2010}
{Konstantinova-Antova} R.,  {Auri{\`e}re} M.,  {Charbonnel} C.,  {Drake} N.~A.,
   {Schr{\"o}der} K.-P.,  {Stateva} I.,  {Alecian} E.,  {Petit} P.,
  {Cabanac} R.,  2010, \aap, 524, A57

\bibitem[\protect\citeauthoryear{{Lazarian} \& {Hoang}}{{Lazarian} \&
  {Hoang}}{2007}]{Lazarian2007}
{Lazarian} A.,  {Hoang} T.,  2007, \mnras, 378, 910

\bibitem[\protect\citeauthoryear{{Lazarian} \& {Hoang}}{{Lazarian} \&
  {Hoang}}{2008}]{Lazarian2008}
{Lazarian} A.,  {Hoang} T.,  2008, \apjl, 676, L25

\bibitem[\protect\citeauthoryear{{Lazarian} \& {Hoang}}{{Lazarian} \&
  {Hoang}}{2011}]{Lazarian2011}
{Lazarian} A.,  {Hoang} T.,  2011, in {Bastien} P.,  {Manset} N.,  {Clemens}
  D.~P.,   {St-Louis} N.,  eds, Astronomical Polarimetry 2008: Science from
  Small to Large Telescopes Vol.~449 of Astronomical Society of the Pacific
  Conference Series, {Alignment of Dust by Radiative Torque: Recent
  Developments}.
p.~116

\bibitem[\protect\citeauthoryear{{Leal-Ferreira}, {Vlemmings}, {Diamond},
  {Kemball}, {Amiri} \& {Desmurs}}{{Leal-Ferreira} et~al.}{2012}]{Leal2012}
{Leal-Ferreira} M.~L.,  {Vlemmings} W.~H.~T.,  {Diamond} P.~J.,  {Kemball} A.,
  {Amiri} N.,    {Desmurs} J.-F.,  2012, \aap, 540, A42

\bibitem[\protect\citeauthoryear{{Leal-Ferreira}, {Vlemmings}, {Kemball} \&
  {Amiri}}{{Leal-Ferreira} et~al.}{2013}]{Ferreira2013}
{Leal-Ferreira} M.~L.,  {Vlemmings} W.~H.~T.,  {Kemball} A.,    {Amiri} N.,
  2013, \aap, 554, A134

\bibitem[\protect\citeauthoryear{{L{\`e}bre}, {Auri{\`e}re}, {Fabas}, {Gillet},
  {Herpin}, {Konstantinova-Antova} \& {Petit}}{{L{\`e}bre}
  et~al.}{2014}]{Lebre2014}
{L{\`e}bre} A.,  {Auri{\`e}re} M.,  {Fabas} N.,  {Gillet} D.,  {Herpin} F.,
  {Konstantinova-Antova} R.,    {Petit} P.,  2014, \aap, 561, A85

\bibitem[\protect\citeauthoryear{{Leone}, {Corradi}, {Mart{\'{\i}}nez
  Gonz{\'a}lez}, {Asensio Ramos} \& {Manso Sainz}}{{Leone}
  et~al.}{2014}]{Leone2014}
{Leone} F.,  {Corradi} R.~L.~M.,  {Mart{\'{\i}}nez Gonz{\'a}lez} M.~J.,
  {Asensio Ramos} A.,    {Manso Sainz} R.,  2014, \aap, 563, A43

\bibitem[\protect\citeauthoryear{{Leone}, {Mart{\'{\i}}nez Gonz{\'a}lez},
  {Corradi}, {Privitera} \& {Manso Sainz}}{{Leone} et~al.}{2011}]{Leone2011}
{Leone} F.,  {Mart{\'{\i}}nez Gonz{\'a}lez} M.~J.,  {Corradi} R.~L.~M.,
  {Privitera} G.,    {Manso Sainz} R.,  2011, \apjl, 731, L33

\bibitem[\protect\citeauthoryear{{McMullin}, {Waters}, {Schiebel}, {Young} \&
  {Golap}}{{McMullin} et~al.}{2007}]{McMullin2007}
{McMullin} J.~P.,  {Waters} B.,  {Schiebel} D.,  {Young} W.,    {Golap} K.,
  2007, in {Shaw} R.~A.,  {Hill} F.,   {Bell} D.~J.,  eds, Astronomical Data
  Analysis Software and Systems XVI Vol.~376 of Astronomical Society of the
  Pacific Conference Series, {CASA Architecture and Applications}.
p.~127

\bibitem[\protect\citeauthoryear{{Ohashi}, {Kataoka}, {Nagai}, {Momose},
  {Muto}, {Hanawa}, {Fukagawa}, {Tsukagoshi}, {Murakawa} \& {Shibai}}{{Ohashi}
  et~al.}{2018}]{Ohashi2018}
{Ohashi} S.,  {Kataoka} A.,  {Nagai} H.,  {Momose} M.,  {Muto} T.,  {Hanawa}
  T.,  {Fukagawa} M.,  {Tsukagoshi} T.,  {Murakawa} K.,    {Shibai} H.,  2018,
  ArXiv e-prints

\bibitem[\protect\citeauthoryear{{Sabin}, {Hull}, {Plambeck}, {Zijlstra},
  {V{\'a}zquez}, {Navarro} \& {Guill{\'e}n}}{{Sabin} et~al.}{2015}]{Sabin2015b}
{Sabin} L.,  {Hull} C.~L.~H.,  {Plambeck} R.~L.,  {Zijlstra} A.~A.,
  {V{\'a}zquez} R.,  {Navarro} S.~G.,    {Guill{\'e}n} P.~F.,  2015, \mnras,
  449, 2368

\bibitem[\protect\citeauthoryear{{Sabin}, {Wade} \& {L{\`e}bre}}{{Sabin}
  et~al.}{2015}]{Sabin2015a}
{Sabin} L.,  {Wade} G.~A.,    {L{\`e}bre} A.,  2015, \mnras, 446, 1988

\bibitem[\protect\citeauthoryear{{Sabin}, {Zhang}, {Zijlstra}, {Patel},
  {V{\'a}zquez}, {Zauderer}, {Contreras} \& {Guill{\'e}n}}{{Sabin}
  et~al.}{2014}]{Sabin2014}
{Sabin} L.,  {Zhang} Q.,  {Zijlstra} A.~A.,  {Patel} N.~A.,  {V{\'a}zquez} R.,
  {Zauderer} B.~A.,  {Contreras} M.~E.,    {Guill{\'e}n} P.~F.,  2014, \mnras,
  438, 1794

\bibitem[\protect\citeauthoryear{{Sahai}, {Morris}, {S{\'a}nchez Contreras} \&
  {Claussen}}{{Sahai} et~al.}{2007}]{Sahai2007}
{Sahai} R.,  {Morris} M.,  {S{\'a}nchez Contreras} C.,    {Claussen} M.,  2007,
  \aj, 134, 2200

\bibitem[\protect\citeauthoryear{{Sahai}, {Morris} \& {Villar}}{{Sahai}
  et~al.}{2011}]{Sahai2011}
{Sahai} R.,  {Morris} M.~R.,    {Villar} G.~G.,  2011, \aj, 141, 134

\bibitem[\protect\citeauthoryear{{S{\'a}nchez Contreras}, {Alcolea},
  {Castro-Carrizo}, {Velilla-Prieto}, {Santander-Garc{\'{\i}}a},
  {Quintana-Lacaci} \& {Cernicharo}}{{S{\'a}nchez Contreras}
  et~al.}{2018}]{SC2018}
{S{\'a}nchez Contreras} C.,  {Alcolea} J.,  {Castro-Carrizo} A.,
  {Velilla-Prieto} L.,  {Santander-Garc{\'{\i}}a} M.,  {Quintana-Lacaci} G.,
  {Cernicharo} J.,  2018, ArXiv e-prints

\bibitem[\protect\citeauthoryear{{Serkowski}}{{Serkowski}}{1974}]{Serkowski74}
{Serkowski} K.,  1974, {Polarization techniques.}.
pp 361--414

\bibitem[\protect\citeauthoryear{{Steffen}, {Hubrig}, {Todt}, {Sch{\"o}ller},
  {Hamann}, {Sandin} \& {Sch{\"o}nberner}}{{Steffen}
  et~al.}{2014}]{Steffen2014}
{Steffen} M.,  {Hubrig} S.,  {Todt} H.,  {Sch{\"o}ller} M.,  {Hamann} W.-R.,
  {Sandin} C.,    {Sch{\"o}nberner} D.,  2014, \aap, 570, A88

\bibitem[\protect\citeauthoryear{{Tazaki}, {Lazarian} \& {Nomura}}{{Tazaki}
  et~al.}{2017}]{Tazaki2017}
{Tazaki} R.,  {Lazarian} A.,    {Nomura} H.,  2017, \apj, 839, 56

\bibitem[\protect\citeauthoryear{{Vlemmings}, {Khouri}, {Mart{\'{\i}}-Vidal},
  {Tafoya}, {Baudry}, {Etoka}, {Humphreys}, {Jones}, {Kemball}, {O'Gorman},
  {P{\'e}rez-S{\'a}nchez} \& {Richards}}{{Vlemmings}
  et~al.}{2017}]{Vlemmings2017}
{Vlemmings} W.~H.~T.,  {Khouri} T.,  {Mart{\'{\i}}-Vidal} I.,  {Tafoya} D.,
  {Baudry} A.,  {Etoka} S.,  {Humphreys} E.~M.~L.,  {Jones} T.~J.,  {Kemball}
  A.,  {O'Gorman} E.,  {P{\'e}rez-S{\'a}nchez} A.~F.,    {Richards} A.~M.~S.,
  2017, \aap, 603, A92

\bibitem[\protect\citeauthoryear{{Vlemmings}, {Ramstedt}, {Rao} \&
  {Maercker}}{{Vlemmings} et~al.}{2012}]{Vlemmings2012}
{Vlemmings} W.~H.~T.,  {Ramstedt} S.,  {Rao} R.,    {Maercker} M.,  2012, \aap,
  540, L3

\bibitem[\protect\citeauthoryear{{Vlemmings} \& {van Langevelde}}{{Vlemmings}
  \& {van Langevelde}}{2008}]{Vlemmings2008}
{Vlemmings} W.~H.~T.,  {van Langevelde} H.~J.,  2008, \aap, 488, 619

\end{thebibliography}

\bsp

\end{document}